\documentclass[aps,prl,10pt,showpacs,floatfix,amsmath,amssymb,notitlepage]{revtex4-1}
\usepackage{amssymb}
\usepackage{bm}
\usepackage{url}
\usepackage{graphicx}
\usepackage{color}
\usepackage[FIGTOPCAP,bf,tight]{subfigure}

\AtBeginDocument{%
    \newwrite\bibnotes
    \def\bibnotesext{Notes.bib}
    \immediate\openout\bibnotes=\jobname\bibnotesext
    \immediate\write\bibnotes{@CONTROL{REVTEX41Control}}
    \immediate\write\bibnotes{@CONTROL{%
    apsrev41Control,author="08",editor="1",pages="1",title="0",year="1"}}
     \if@filesw
     \immediate\write\@auxout{\string\citation{apsrev41Control}}%
    \fi
}%

\begin{document}






\title{Fluid dynamics of COVID-19 airborne infection suggests urgent data for a scientific design of social distancing}





\author{M.~E. Rosti$^{1}$}
\email{marco.rosti@oist.jp}
\author{S. Olivieri$^{1}$}
\author{M. Cavaiola$^{2,3}$}
\author{A. Seminara$^{4}$}
\author{A. Mazzino$^{2,3}$}
\email{andrea.mazzino@unige.it}

\affiliation{
$^1$ Complex Fluids and Flows Unit, Okinawa Institute of Science and Technology Graduate University, 1919-1 Tancha, Onna-son, Okinawa 904-0495, Japan\\ 
$^2$ Department of Civil, Chemical and Environmental Engineering (DICCA), University of Genova, Via Montallegro 1, 16145, Genova, Italy \\
$^3$ INFN, Genova Section, Via Montallegro 1, 16145, Genova, Italy \\
$^4$ CNRS and Universit\'e C\^ote d'Azur, Institut de Physique de Nice, UMR7010, 06108 Nice, France
}

\begin{abstract}
\section{Abstract} \label{sec:abstract}
The COVID-19 pandemic is largely caused by airborne transmission, a phenomenon that rapidly gained the attention of the scientific community. Social distancing is of paramount importance to limit the spread of the disease, but to design social distancing rules on a scientific basis the process of dispersal of virus-containing respiratory droplets must be understood. Here, we demonstrate that available knowledge is largely inadequate to make predictions on the reach of infectious droplets emitted during a cough and on their infectious potential. We follow the position and evaporation of thousands of respiratory droplets by massive state-of-the-art numerical simulations of the airflow caused by a typical cough. We find that different initial distributions of droplet size taken from literature and different ambient relative humidity lead to opposite conclusions: (1) most \emph{vs} none of the viral content settles in the first 1-2 m; (2) viruses are carried entirely on dry nuclei \emph{vs} on liquid droplets; (3) small droplets travel less than $2.5\,\mathrm{m}$ \emph{vs} more than $7.5\,\mathrm{m}$. We point to two key issues that need to be addressed urgently in order to provide a scientific foundation to social distancing rules: (I1) a careful characterisation of the initial distribution of droplet sizes; (I2) the infectious potential of viruses carried on dry nuclei \emph{vs} liquid droplets.

\end{abstract}


\maketitle 

\section{Introduction} \label{sec:introduction}
The airborne transmission route of SARS-CoV-2 certainly deserves the numerous ongoing efforts aimed at fighting the COVID-19 pandemic~\cite{bahl2020airborne,bourouiba2020turbulent,mittal2020flow,lewis2020coronavirus,lincei2020review}. It is well known that SARS-CoV-2 infection relies on the spreading of small virus-containing respiratory droplets that the infected person exhales when coughing or sneezing or even simply talking or breathing~\cite{asadi2019aerosol}. However, at least two unresolved key issues (I1 and I2 in the following) remain open~\cite{morawska2020time,lincei2020review} and need urgent attention.

First (I1): we need to better characterize the sizes of the exhaled droplets for all the expulsion processes, coughing, speaking, breathing and sneezing~\cite{lincei2020review,mittal2020flow}. \citet{flugge1897ueber} and~\citet{wells1934air} have highlighted the importance of this issue. \citet{wells1934air} and~\citet{duguid1945numbers,duguid1946size} were the first to propose systematic measurements of droplet sizes.
After their seminal papers, many investigators
have grappled with issue I1 (see e.g.~\cite{papineni1997size,morawska2009size,xie2009exhaled,yang2007size,johnson2011modality,chao2009characterization,zayas2012cough,han2013characterizations}, among others).   
A careful analysis of the state of the art on the subject reported in \citet{lincei2020review} shows broad differences in the experimental results of the different investigators. For example, \citet{zayas2012cough} state that the droplets in the sub-micron range represent 97$\%$ of the exhaled droplets for each single cough event; for the same type of expulsion, \citet{yang2007size} report a much smaller percentage of less than 4$\%$ while not even a single droplet within this subrange was measured by~\citet{duguid1946size}.
On the one hand, the physics underpinning the formation of respiratory droplets is not completely understood~\cite{lincei2020review}. On the other hand, experiments exploit different techniques under different ambient conditions. Finally, a rigorous presentation of data is not always provided.
This lack of a systematic analysis, in addition to the natural variability across individuals, may explain the striking inconsistency of available information on the size distribution of exhaled droplets.  

 Second (I2): we need to establish whether viruses lingering on dry nuclei upon droplet evaporation retain their full potential of infection. 
There is evidence supporting that viruses coated by a lipid membrane tend to retain their infectivity longer at low relative humidity \cite{sobsey2003virus}. The coated SARS-CoV-2 virus is thus expected to best thrive in dry conditions~\cite{lincei2020review}. However, the opposite is true in relevant counterexamples as discussed by \citet{yang2012mechanisms}.

The two issues listed above cause considerable uncertainty in the expected efficiency of disease transmission. This uncertainty stems from a rather simple concept: smaller liquid droplets are lighter, hence remain airborne for longer times and are more likely to shrink to their dry residual nuclei under sufficiently dry ambient conditions. Hence the infection potential of a single cough or sneeze depends critically on the size distribution of exhaled droplets and the likelihood of disease transmission through viruses carried on dry nuclei \emph{vs} liquid droplets.

\begin{figure}[h!]
\centering
\begin{minipage}{0.35\textwidth}
\centering
\subfigure[][]{
\includegraphics[width=1.2\textwidth]{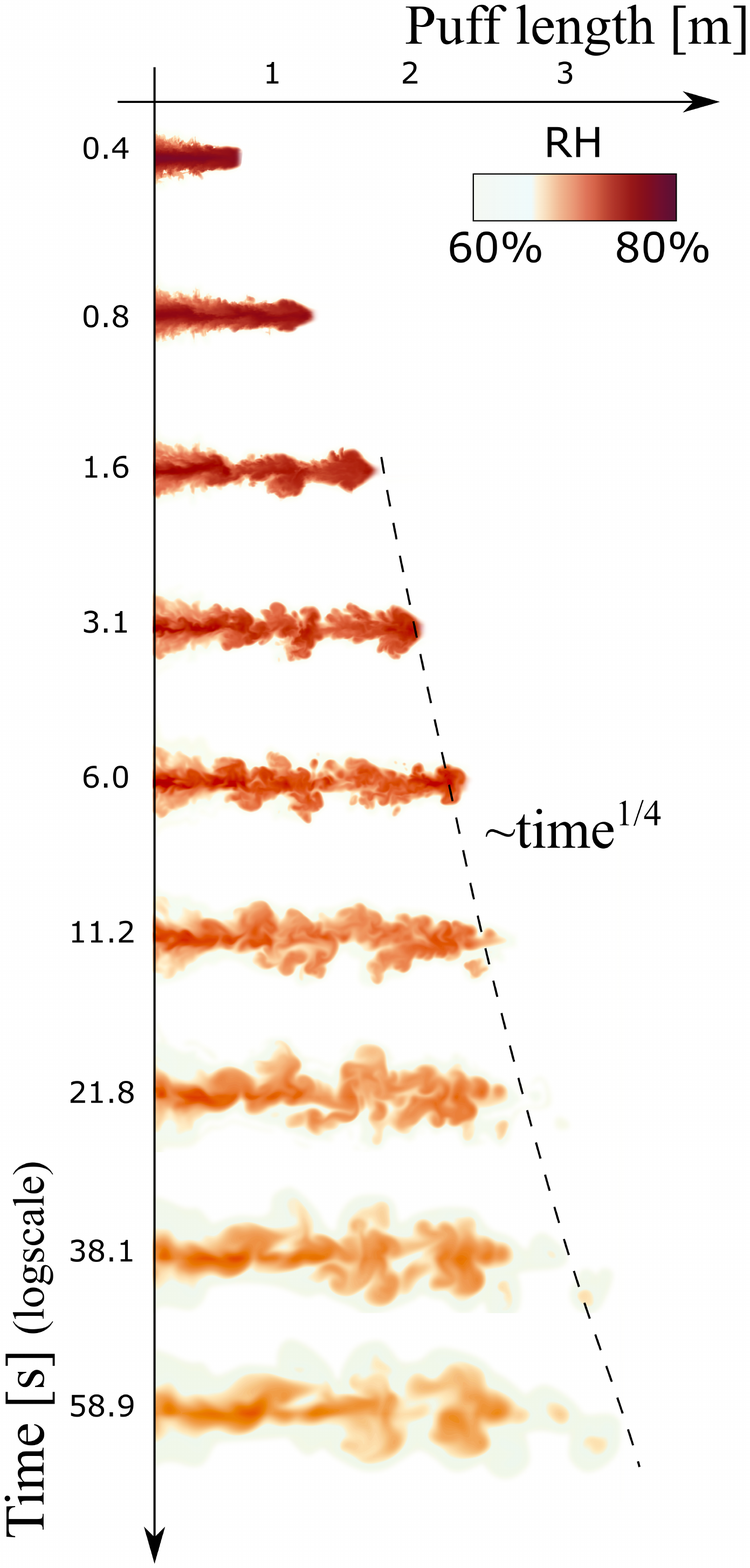}}
\end{minipage}\qquad
\begin{minipage}{0.6\textwidth}
\centering
\subfigure[][]{
\includegraphics[width=0.7\textwidth]{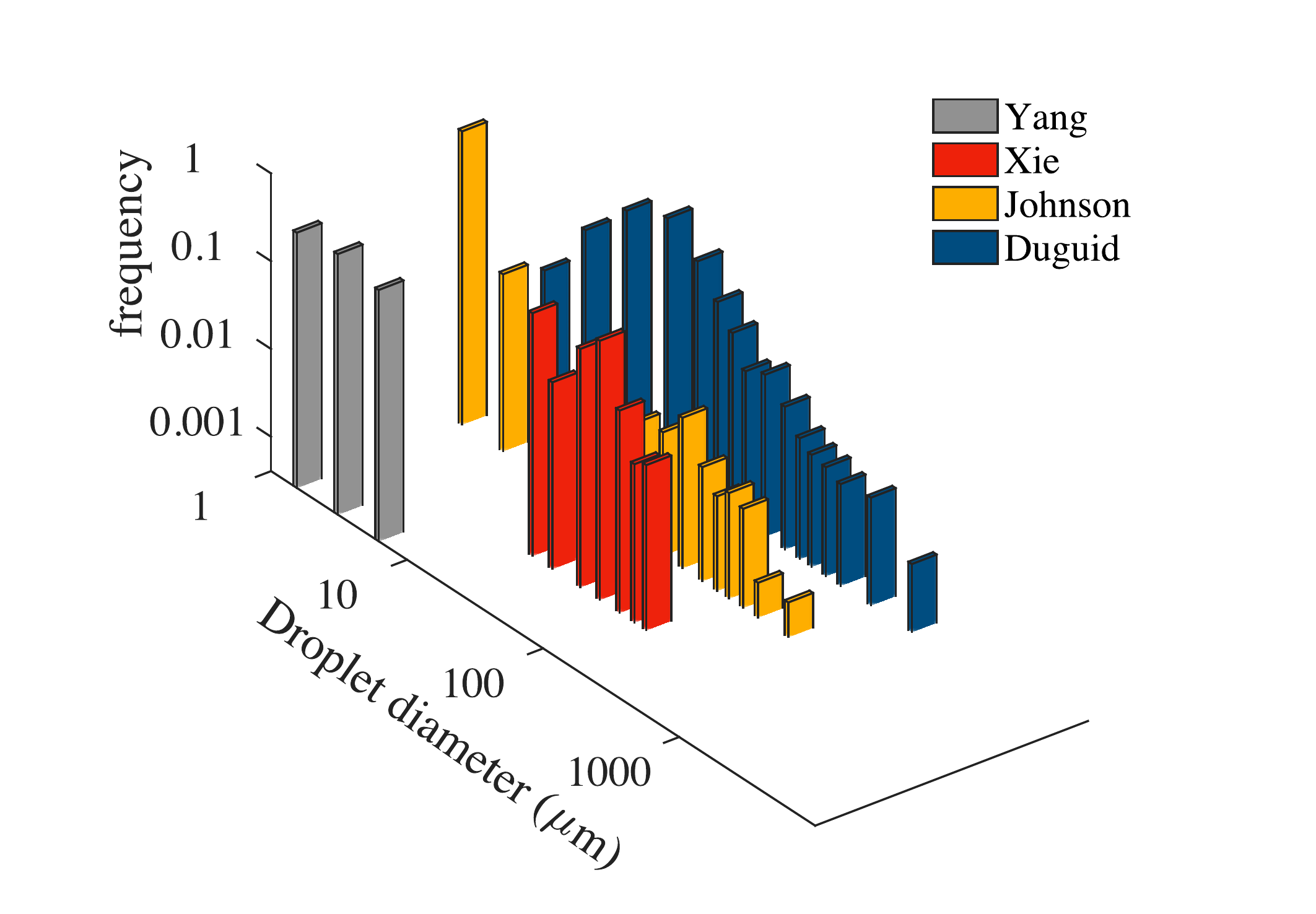}}\\[10pt]
\subfigure[][]{
\includegraphics[width=.9\textwidth]{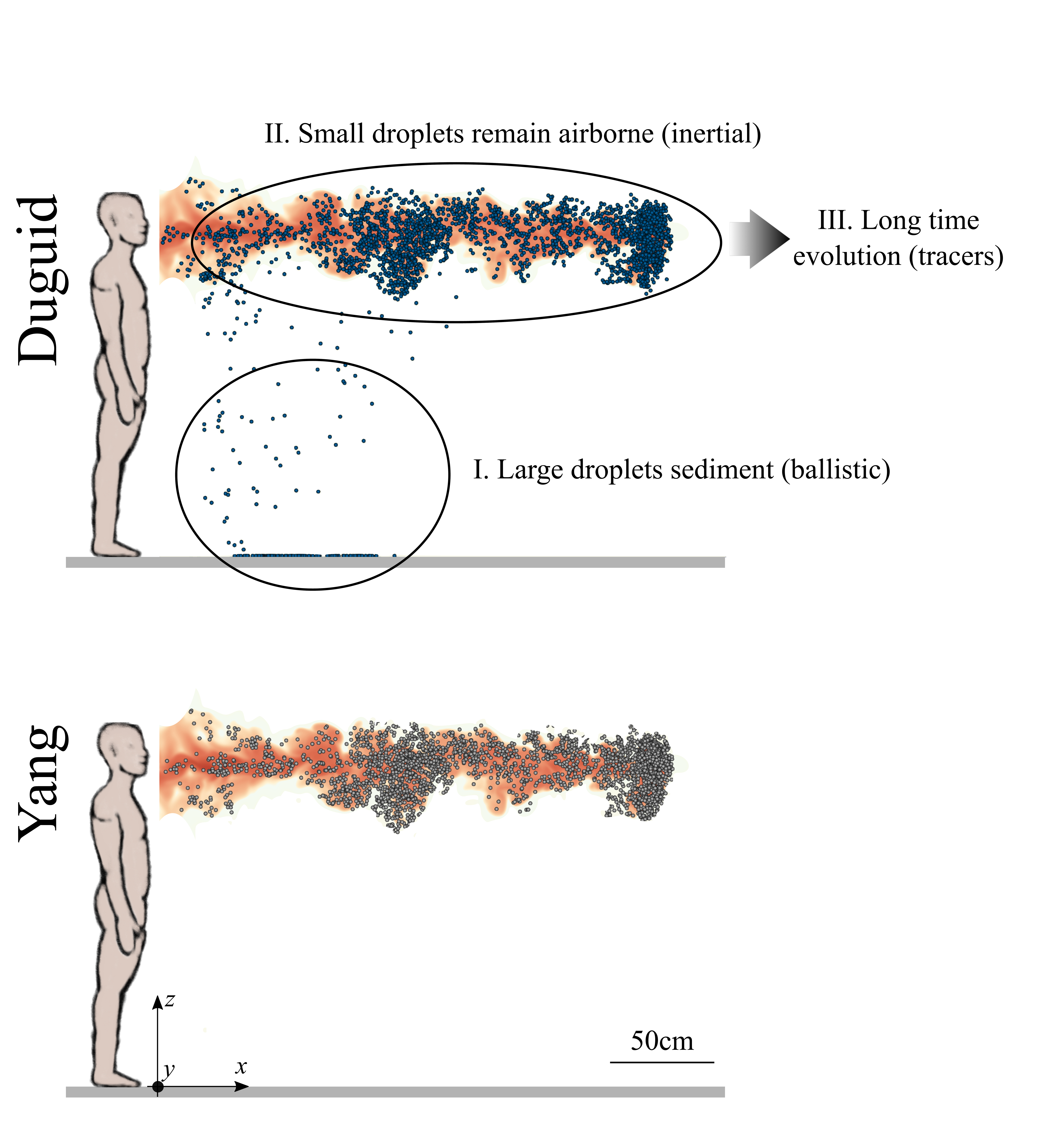}}
\end{minipage}
\caption{\textbf{Airflow generated during coughing.}
\textbf{a} Evolution of the relative humidity in space and time. 
After the end of the exhalation, the emitted air behaves as a turbulent puff growing in length as $\sim t^{1/4}$ and decaying in amplitude as $\sim t^{-3/4}$ (the latter is shown in Fig~\ref{fig:valPuff}.).
\textbf{b} Droplet initial size distributions considered in the present study: \citet{duguid1946size} (blue), \citet{johnson2011modality} (yellow), \citet{xie2009exhaled} (red), \citet{yang2007size} (gray).
\textbf{c} Relative humidity (color coded) and exhaled droplets (blue and gray spheres, not in scale) after $7.6\,\mathrm{s}$ considering two different initial droplet size distributions: (top) \citet{duguid1946size}; (bottom) \citet{yang2007size} showcases the dramatic differences in predictions depending on the initial distribution of droplet sizes. The distribution of droplet sizes from \citet{duguid1946size} contains large droplets that rapidly settle carrying most viral load on the ground, as well as many small droplets which remain airborne. In contrast, in the size distribution from \citet{yang2007size} all droplets are small enough to remain airborne for the entire simulation. The ambient RH is $60 \%$ in all figures. Scale bar: 50 cm.
 }
\label{fig:sketch}
\end{figure}

But to what extent do predictions actually vary across different scenarios for I1 and I2? To answer this question we quantify the viral load carried on dry nuclei \emph{vs} liquid droplets upon cough. We leverage concepts developed in the context of atmospheric cloud formation~\cite{pruppacher1997microphysics} to track the evaporation of respiratory droplets as they move away from the mouth (Fig.~\ref{fig:sketch}). To simultaneously monitor droplet position and evaporation we employ massive state-of-the-art direct numerical simulations (DNS) of the airflow and humidity (see Methods section for details). DNS are the most powerful and demanding tools of computational fluid dynamics and are mandatory to fully capture the key role of turbulence on the spreading of virus-containing droplets, as recently shown in Refs.~\cite{chong2020extended,rosti2020turbulence}. 
We conduct a systematic comparative analysis across eight scenarios selected from the literature~\cite{duguid1946size,johnson2011modality,xie2009exhaled,yang2007size} and demonstrate that different initial conditions yield entirely different conclusions. 
Depending on the distribution of droplet initial size, \emph{(i)} most \emph{vs} none of the viral content settles in the first 1-2 m; \emph{(ii)} all viruses are carried in the air on dry nuclei \emph{vs} on liquid droplets; \emph{(iii)} small droplets settle slowly on the ground and travel less than $2.5\,\mathrm{m}$ \emph{vs} more than $7.5\,\mathrm{m}$. We focus on the airflow generated by the cough: further work will focus on the effects of aeration in the environment, especially at long times.

\section{Results} \label{sec:results}

We simulate a strong expiratory event typical of cough~\cite{gupta2009flow} causing an unsteady jet of humid air that evolves into a turbulent puff while becoming drier and carries many virus-containing droplets (Fig.~\ref{fig:sketch}). 
We record the position of each single droplet inside the turbulent cloud of exhaled air, while simultaneously monitoring their liquid content. Keeping track of the entire ensemble is crucial to quantify systematically the amount of viral load carried on dry nuclei \emph{vs} liquid droplets in space and time, which dictates the associated risk of transmission on the basis of issue I2. \\

We conduct eight numerical experiments considering two different levels of ambient relative humidity (RH=60$\%$ prefix `Wet' and RH=40\% prefix `Dry') combined with four different initial size distributions of the exhaled droplets.  
Despite being relatively similar in terms of water content, our `Wet' and `Dry' conditions illustrate a key effect, which often goes unnoticed. Our `Wet' condition lays \emph{above} the efflorescence RH, namely droplets never evaporate completely but remain in the liquid state in equilibrium with the surrounding ambient humidity. Conversely, our `Dry' condition is \emph{below} the efflorescence RH hence all droplets eventually evaporate completely and shrink to their dry nuclei~\cite{lohmann2016introduction}. 
We simulate the `Dry' and `Wet' conditions for four different droplet size distributions (Fig.~\ref{fig:sketch}b) proposed by \citet{duguid1946size} (suffix `Du'), by~\citet{johnson2011modality}
(suffix `Jo'), by~\citet{xie2009exhaled} (suffix `Xi'), and by~\citet{yang2007size} (suffix `Ya').
The eight experiments are labelled: WetDu, WetJo, WetXi, WetYa, and similarly for the `Dry' condition.

A snapshot of droplet positions demonstrates the undeniable role of droplet size at emission (Fig.~\ref{fig:sketch}c). 
The distribution of droplet sizes from~\citet{duguid1946size} (Fig.~\ref{fig:sketch}c top) yields a scenario largely consistent with the literature~\cite{mittal2020flow}, where droplets belong to either of two classes. Large droplets sediment owing to their weight with negligible action of the airflow (phase I in Fig.~\ref{fig:sketch}c top); small droplets remain airborne and travel within the turbulent puff (phase II in Fig.~\ref{fig:sketch}c top); after few seconds they reach their minimum size and they are carried as tracers by the airflow (phase III in Fig.~\ref{fig:sketch}c top). But a different distribution of droplet sizes, \citet{yang2007size}, yields an entirely different picture (Fig.~\ref{fig:sketch}c bottom): there are no large droplets, and the entire viral load is carried on small airborne droplets that never settle in our simulation.

\subsection{I. Loss of viral load via sedimentation to the ground (ballistic)}

To quantify these observations we define the (relative) viral load of the $i$-th droplet as the ratio between its initial volume and the cumulative initial volume of all exhaled droplets. In other words, we assume that the viral load of a given droplet is proportional to its initial volume and when the droplet undergoes evaporation the viral load is  conserved (i.e. any degradation of the virus is neglected). This assumption is sensible in view of the recent findings by~\citet{fears2020persistence} showing that the SARS-CoV-2 virus retains infectivity and integrity up to $16$ hours in laboratory-created respirable-sized aerosols. 
The model may be extended to account for variations of viral load depending on more specific details of the SARS-CoV-2 infection that are currently unknown (e.g.~the likelihood of infecting different parts of the respiratory tract).
\begin{figure}[h!]
\centering
\subfigure[][]{
\includegraphics[width=0.4\textwidth]{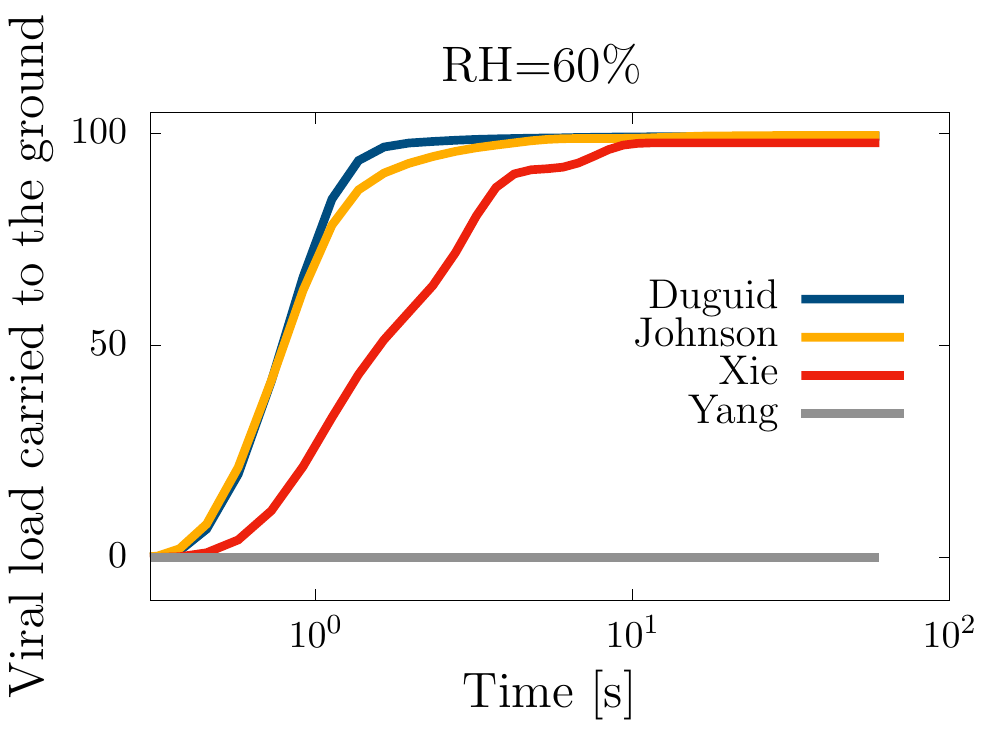}}
\subfigure[][]{
\includegraphics[width=0.4\textwidth]{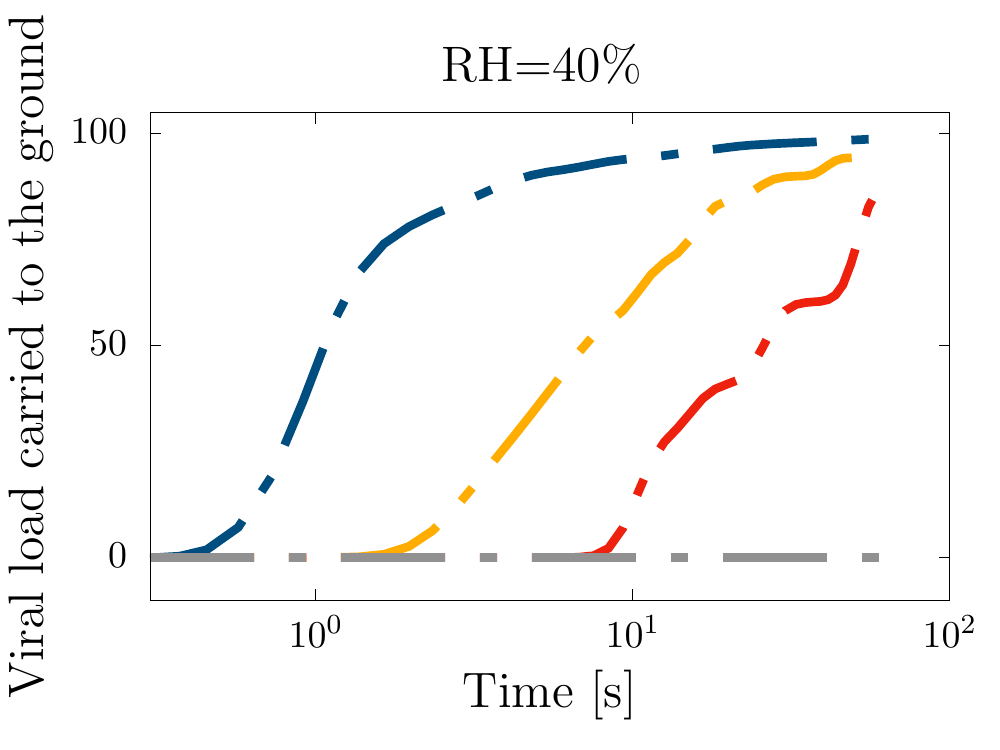}}
\subfigure[][]{
\includegraphics[width=0.85\textwidth]{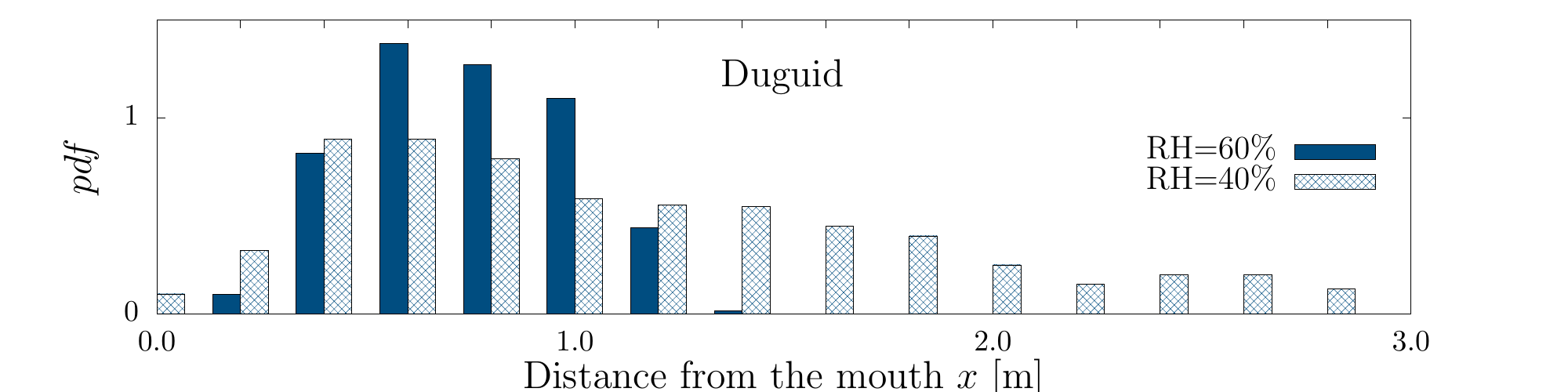}}
\caption{\textbf{Sedimentation of large droplets.}
\textbf{a} Cumulative viral load sedimenting to the ground, obtained with the four different initial droplet size distributions proposed by~\citet{duguid1946size} (blue),~\citet{johnson2011modality} (yellow),~\citet{xie2009exhaled} (red) and~\citet{yang2007size} (gray). Here, the ambient relative humidity is RH=60$\%$. 
\textbf{b} Same as (a), for a dry environment, RH=40$\%$.
Most of the viral load settles within $60\,\mathrm{s}$ for three initial distributions, whereas for one,~\citet{yang2007size}, no droplets settle within the simulation time. \textbf{c} Probability density function of the distance from the mouth when droplets reach the ground; ambient relative humidity $\mathrm{RH}=60\%$ (solid blue) and $\mathrm{RH}=40\%$ (patterned blue). Drier environments cause further spreading: Droplets that reach the ground remain within $1\,\mathrm{m}$ from the mouth in wet conditions, whereas they can reach nearly $3\,\mathrm{m}$ in dry conditions.
}
\label{fig:sed}
\end{figure}

In the first few seconds after exhalation, the puff rapidly loses viral load carried by larger droplets that reach the ground owing to their own weight. 
The amount of viral load lost through sedimentation depends dramatically on the ambient humidity and the initial distribution of droplet sizes (issue I1). For three initial conditions, nearly the entire viral load is carried to the ground after 1 to 3 seconds ($99\%$ for the Du and Jo distributions and $45\%$ for Xi); whereas for the last condition (Ya), exactly zero viral load reaches ground for the entire simulation (Fig.~\ref{fig:sed}a,b).
The inconsistency among predictions for the four size distributions is even more pronounced in the dry cases reported in Fig.~\ref{fig:sed}b (Du:~94$\%$;  Jo:~$61\%$; Xi:~12$\%$; Ya:~0$\%$).
A summary of the cumulative viral load sedimenting to the ground after the entire simulation ($60\,\mathrm{s}$)  is reported in Tab.~\ref{tab:airborne} (VL -- sed). \\ The table additionally shows a different observable, often discussed in the literature: the number of sedimenting droplets normalized to the total number of droplets (ND -- sed). Using this variable can be extremely misleading in the presence of very large droplets as these may be a negligible fraction of the total number of droplets but nonetheless carry nearly the entire viral load owing to their large volume. This is the case for the Du and Jo distributions, for which most viral load settles to the ground carried by few large droplets, yet more than 90\% of the droplets are small and still remain aloft (see Fig.~\ref{fig:ext-sed}a,b).

To complement this analysis, Fig.~\ref{fig:sed}c shows the normalized histogram (probability density function or $pdf$) of the distance travelled by these large droplets when they reach ground, comparing the two ambient conditions for one size distribution (Du). The effect of the ambient humidity is clearly noticeable, with large droplets settling within $1\,\mathrm{m}$ in the Wet condition \emph{vs} almost $3\,\mathrm{m}$ in the Dry condition. Similar results hold for the other size distributions except for Ya for which all droplets remain airborne (see Fig.~\ref{fig:ext-sed}c).

\begin{table}
\centering
\setlength{\tabcolsep}{5pt}
\begin{tabular}{c|ccccccccc}
	& WetDu	& DryDu	& WetJo	& DryJo	& WetXi	& DryXi	& WetYa	& DryYa	\\
\hline
ND -- sed (\%)	& $6$			& $3$		& $5$			& $3$		& $71$		& $45$		& $0$	& $0$		\\
\hline
ND -- 1m (\%)	& $85$			& $90$		& $87$			& $91$		& $24$		& $86$		& $93$	& $93$	\\
ND -- 2m (\%)	& $45$			& $47$		& $48$			& $50$		& $8$		& $32$		& $51$	& $50$	\\
ND -- 4m (\%)	& $8$			& $9$		& $9$			& $10$		& $1$		& $5$		& $10$	& $11$	\\
\hline
ND -- 1m, small (\%)	& $85$			& $89$		& $86$			& $89$		& $19$		& $46$		& $93$	& $93$	\\
ND -- 2m, small (\%)	& $45$			& $46$		& $48$			& $49$		& $8$		& $21$		& $51$	& $50$	\\
ND -- 4m, small	 (\%)& $8$			& $9$		& $9$			& $10$		& $1$		& $4$		& $10$	& $11$	\\
\hline
VL -- sed (\%)	& $99$			& $99$		& $99$			& $95$		& $99$		& $88$		& $0$	& $0$		\\
\hline
VL -- 1m (\%)	& $5$			& $3$		& $26$			& $15$		& $26$		& $60$		& $92$	& $92$	\\
VL -- 2m (\%)	& $0.1$			& $0.6$		& $0.01$		& $2$		& $0.2$		& $13$		& $51$	& $49$	\\
VL -- 4m (\%)	& $0.02$		& $0.07$	& $0.0001$	& $0.08$	& $0.03$	& $1$		& $10$	& $10$	\\
\hline
VL -- 1m, small (\%)	& $0.3$			& $1$		& $0.08$		& $3$		& $0.6$		& $7$		& $92$	& $92$	\\
VL -- 2m, small	(\%) & $0.1$			& $0.4$	& $0.01$		& $1$		& $0.2$		& $3$		& $51$	& $49$	\\
VL -- 4m, small (\%) & $0.02$		& $0.05$	& $0.0001$	& $0.08$	& $0.03$	& $0.6$		& $10$	& $10$	\\
\hline
\end{tabular}
\caption{The cumulative number of droplets (ND) and viral load (VL) measured in the numerical experiments. Note that all the values are given in percentage. Quantities denoted with `sed' correspond to droplets that settle on the ground within the simulation; Quantities denoted with `1m', `2m' and `4m' correspond to airborne droplets travelling up to distances of 1, 2 and 4 metres respectively; Quantities denoted with `small' correspond to to droplets with diameter smaller than $10\,\mathrm{\mu m}$. 
}
\label{tab:airborne}
\end{table}

\begin{figure}
\centering
\subfigure[a][]{
\includegraphics[height=0.33\textwidth]{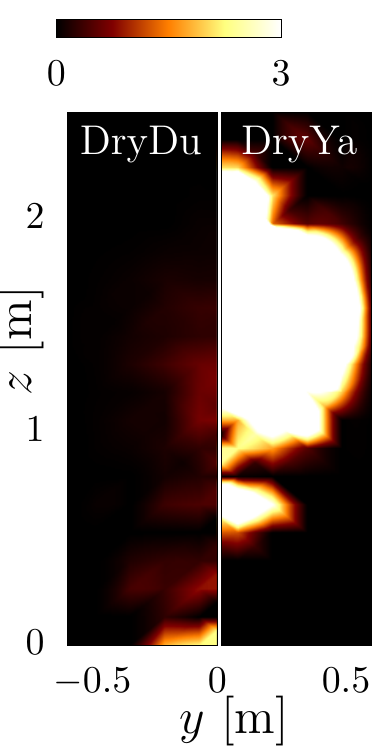}}
\subfigure[b][]{
\includegraphics[height=0.33\textwidth]{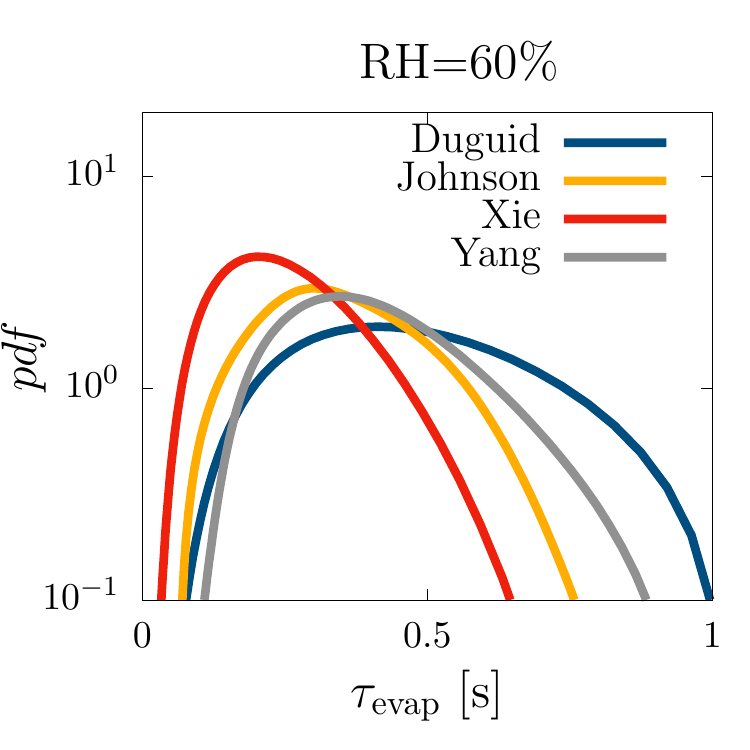}}
\subfigure[c][]{
\includegraphics[height=0.33\textwidth]{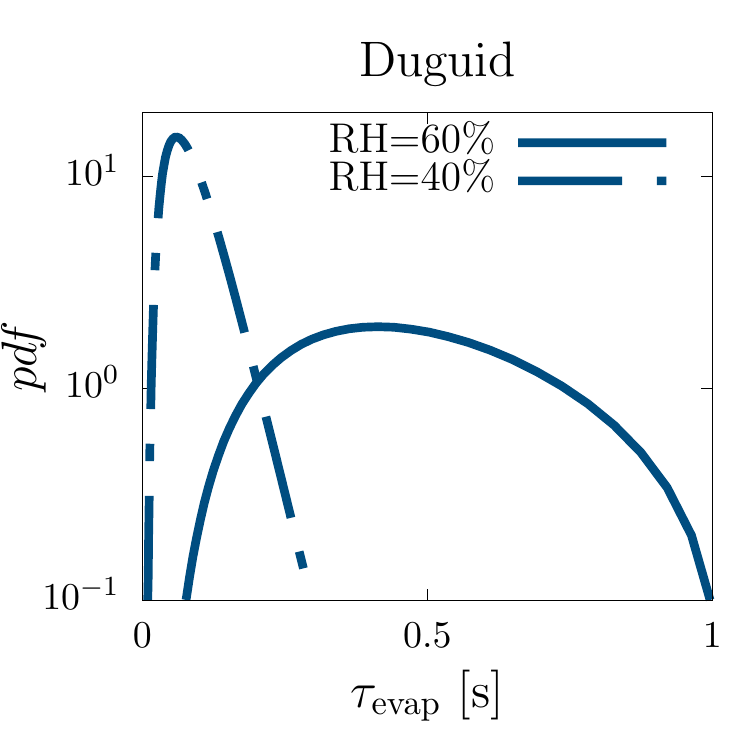}} 
\subfigure[a2][]{
\includegraphics[height=0.33\textwidth]{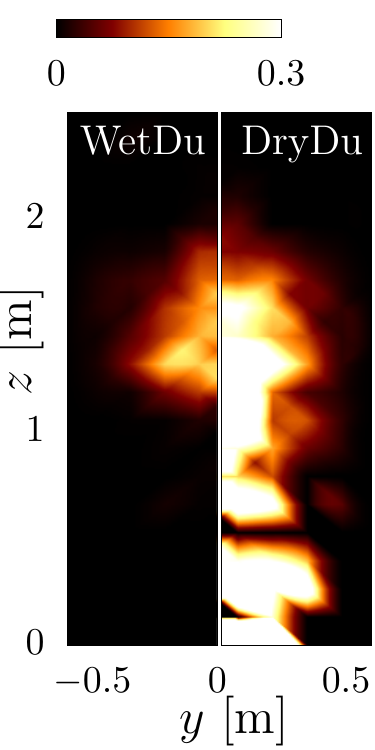}}
\subfigure[d][]{
\includegraphics[height=0.33\textwidth]{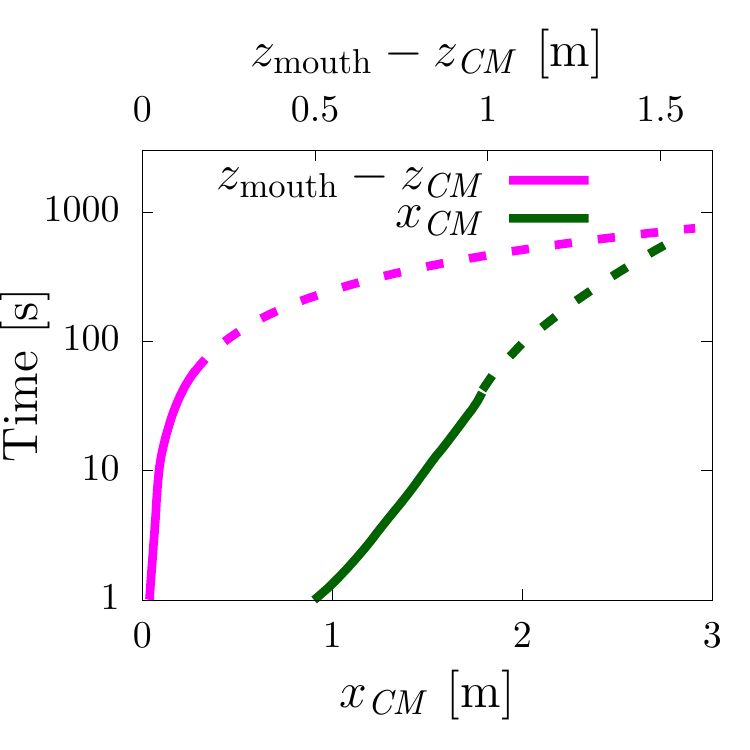}}
\subfigure[e][]{
\includegraphics[height=0.33\textwidth]{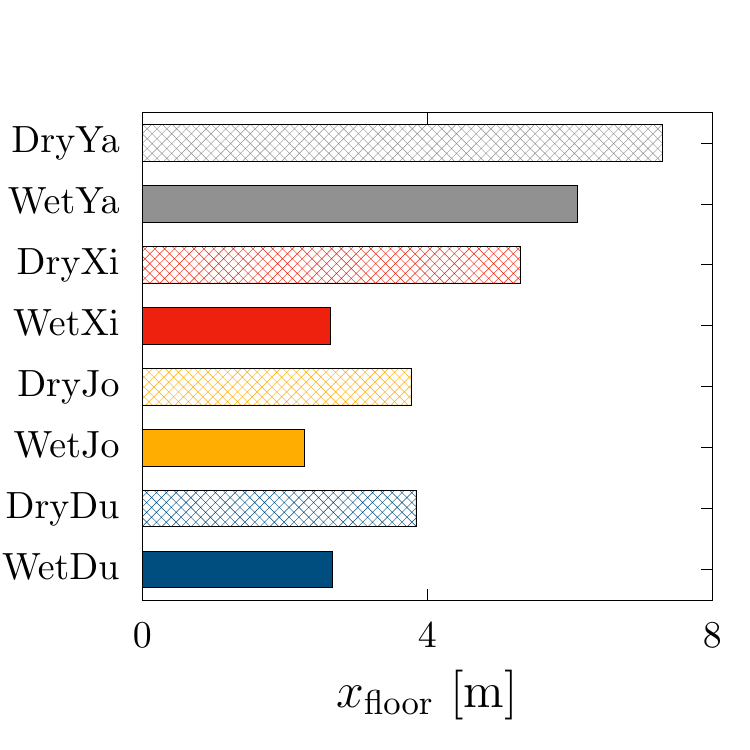}}
\caption{\textbf{Airborne-transmitted droplets.}
\textbf{a,d} Cumulative viral load per unit area (\% viral load$/ \mathrm{m}^2$) reaching a distance of $2\,\mathrm{m}$ from the mouth after $60\,\mathrm{s}$. Results obtained with RH=40$\%$ using the distribution by~\citet{duguid1946size}  and~\citet{yang2007size} (\textbf{a} left and right respectively) and using the distribution by~\citet{duguid1946size} with RH=60$\%$ and 40$\%$ (\textbf{d} left and right respectively).
\textbf{b,c} Probability density function of droplet evaporation time (i.e.~time for the droplet to shrink to its final radius; only airborne droplets in the observation time of $60\,\mathrm{s}$ are considered).
\textbf{b}: results with ambient RH=$60\%$ for the four different initial droplet size distributions, i.e.~\citet{duguid1946size} (blue),~\citet{johnson2011modality} (yellow),~\citet{xie2009exhaled} (red) and~\citet{yang2007size} (gray). \textbf{c}, results for the distribution by~\citet{duguid1946size} with ambient RH=$60\%$ (solid) and RH=$40\%$ (dashed).
The initial size distribution and the ambient humidity cause dramatic differences in the reach of airborne droplets, with variations of the order of $80\%$ for the mean value.
\textbf{e} Trajectory of the viral load center of mass (computed considering only the airborne droplets and not those that already settled on the ground) for the simulation labeled WetDu; horizontal position $x_\mathit{CM}$ (green) and vertical position $z_\mathit{CM}$ (magenta).
The solid lines indicate the results from the simulation while the dashed ones are extrapolations over longer time as discussed in the Methods section.
\textbf{f} Extrapolated horizontal distance travelled by the viral load center of mass for the eight numerical experiments performed.
}
\label{fig:air}
\end{figure}

\subsection{II. Transport of airborne droplets (inertial)}

The fate of smaller droplets is dictated by the interplay between their inertia and the airflow, and thus it depends critically on droplet initial size and subsequent evaporation. Once again we find radically different predictions for the viral load carried by airborne droplets depending on the ambient relative humidity and initial droplet size distribution. The discrepancy in the predictions can be appreciated qualitatively in Fig.~\ref{fig:air}a,d where we map the cumulative viral load per unit area that travel across a vertical plane at $2\,\mathrm{m}$ from the mouth in the entire simulation. 
In the DryYa condition, considerable viral load reaches beyond $2\,\mathrm{m}$ from the mouth in 60 seconds (see Fig.~\ref{fig:air}a and Table~\ref{tab:airborne}, total 49\%), whereas nearly no viral load travels the same distance in the DryDu condition (total 0.6\%). Fig.~\ref{fig:air}d showcases the dramatic effect of the ambient humidity for one initial droplet size distribution (Du).

Synthetic data are summarized in  Tab.~\ref{tab:airborne}, where we report the cumulative viral load carried by airborne droplets reaching a distance of $1\,\mathrm{m}$ (indicated as `VL -- 1m'), $2\,\mathrm{m}$ (`VL -- 2m') and $4\,\mathrm{m}$ (`VL -- 4m') from the mouth within the total observation time of $60\,\mathrm{s}$.
Predictions vary dramatically depending on the relative humidity and the initial droplet size distribution. E.g.~as much as 10\% (DryYa) or as little as 10$^{-4}$\% (WetJo) of the viral load travels 4~m or more from the mouth in 60 seconds. 
Importantly, similar uncertainties persist also when considering only droplets that are smaller than  $10\, \mathrm{\mu m}$ (see results labeled `VL -- 1m, small', `VL -- 2m, small' and `VL -- 4m, small' in Tab.~\ref{tab:airborne}). These droplets are candidate to reach pulmonary alveoli causing the most severe complications of COVID-19~\cite{hinds1999aerosol,nicas2005toward} and their initial volume affects predictions of airborne infection risk models (see e.g.~the model proposed by  \citet{nicas2005toward} and its Eq.~1). Hence the uncertainty in the initial droplet size distribution (I1) affects dramatically the reliability of airborne infection risk models.

\subsection{III. Long-range transmission (tracers)}

How does the journey of these airborne droplets proceed after the end of our simulation?
After few seconds, all droplets are either liquid at their final equilibrium size (RH$=60\%$) or shrinked to their dry nucleus (RH$=40\%$); either way, they behave as fluid tracers. Their final destination depends on the external airflow hence on the specific indoor or outdoor environment and its aeration. 
In order to provide a simple estimate for the ultimate reach of the viral load, we ignore the presence of external airflow and we track the center of mass of the airborne viral load in time excluding the sedimenting droplets (see Fig.~\ref{fig:air}e for a typical trace). We extrapolate the trajectory to the location where the center of mass eventually reaches the ground (see Fig.~\ref{fig:air}f and Methods for details). This simple estimate shows that even in the absence of external airflow, small droplets travel several meters. Once again we observe a remarkable variability: while for WetJo the spreading is contained in less than $2.5\,\mathrm{m}$, for DryYa droplets travel beyond $7.5\,\mathrm{m}$. Airborne droplets reach the floor in about $20\,\mathrm{min}$ which is well within the 16 hours of virus survival recently measured by~\citet{fears2020persistence}. Note that in the Dry condition, the viral load reaches the floor on dry nuclei because droplets fully evaporate, whereas in the Wet condition viruses travel on droplets that retain their liquid content. Once again, the two issues I1 and I2 are crucial to establish the reach and infectious potential of droplets expelled in a cough. \\

Another important observable is the concentration of infectious material, which is inversely proportional to the volume of the cloud of droplets. The cloud expands at a rate that is intertwined with the turbulent nature of the cough. When droplets are shrunk to their final size and closely follow the airflow, the size of the cloud grows as $t^{1/4}$ (see Methods and Fig.~\ref{fig:valPuff}b).
This scaling holds for all our simulations at long times, as it is a fundamental property of the turbulent airflow generated by the cough. However, prior to this regime, most droplets are inertial and they follow the flow with delays that depend on their size. Hence the cloud of droplets expands at a rate that depends on droplet size distribution and evaporation. This regime is extremely complex and requires an in depth description of turbulent fluctuations. Indeed, the interaction between inertial effects and turbulence causes nontrivial correlations and ultimately slows down evaporation~\cite{rosti2020turbulence}. The distribution of evaporation times resulting from these non-trivial effects varies considerably across different conditions (Fig.~\ref{fig:air}b,c). 
In our simulations, the droplet cloud expands at different rates depending on the initial condition (see Fig.~\ref{fig:sigma2}). Although variations are sizeable (about 30\%), they are overshadowed by the much more dramatic variations in the position of the center of mass (100\%).

Our results are obtained in the absence of external airflows: in our simulations air is set in motion by the cough only. Because cough is a violent respiratory event, environmental airflow is negligible in the first phases of the dynamics. However, in the long time regime, the direct effect of the cough fades and external airflows dictate the final reach of small respiratory droplets. In order to provide more specific predictions about the final reach of small droplets, our approach must be adapted to a particular environment and account for its specific airflow and aeration. 

\section{Discussion} \label{sec:discussion}

Current guidelines released by WHO for the protection from airborne virus transmission introduce the notion of a safe distance of 1-2 meter to ensure protection from an infected individual.
In the present paper we discussed the scientific foundation of these  `social distancing' measures, which touches several billion individuals globally. 
Our simulations demonstrate that currently available information is inadequate to design social distancing recommendations on a solid scientific basis.
Indeed, diametrically opposed predictions are drawn depending on the size distribution of the respiratory droplets and ambient humidity: \emph{(i)} most \emph{vs} none of the viral load settles in the first 1-2 m in few seconds; \emph{(ii)}  all viral load is carried on dry nuclei \emph{vs} liquid droplets and \emph{(iii)} small airborne particles travel less than $2.5\,\mathrm{m}$ \emph{vs} more than $7.5\,\mathrm{m}$. 
Our findings call for novel experimental efforts to address two key issues that cause uncertainty in predictions: the determination of droplet size distributions at emission (I1) and the infection potential of viral load carried on dry \emph{vs} wet nuclei (I2).

Our central observable here is the relative viral load, i.e.~the amount of virus carried by an individual droplet normalized to the total amount of virus in the ensemble of droplets. To connect the relative viral load to the probability of infection, further information is needed: what is the cumulative viral load emitted with the entire population of droplets by an infected individual, and what is the infectious dose of SARS-CoV-2? A comprehensive revision of the state of knowledge on this complex issue is beyond the scope of the present work. However we note that while epidemiology and virology are clearly at the front line in the fight against COVID-19, knowledge on the disease itself must be coupled to the physics of droplet production, transport and evaporation.
We hope that our work will raise awareness about these less appreciated unknowns of the problem.

Finally, our results show that a single rule for social distancing may not be adequate to protect individuals in different environments. The relative humidity of the environment has a particularly dramatic effect, with all droplets evaporating to their dry nuclei under sufficiently dry conditions, and all droplets remaining liquid under sufficiently wet environmental conditions. Provided science advances on the key issues identified above, the strategy employed in the current study can actively contribute to outline a revised notion of social distancing underpinned by scientific evidence. 


%


\clearpage

\section{Methods} \label{sec:methods}
\subsection{Direct numerical simulations of cough-generated airflow}
The airflow is governed by the incompressible Navier--Stokes equations
\begin{equation}
  \partial_t \bm{u} + \bm{u}\cdot \bm{\partial} \bm{u}=-\frac{1}{\rho_a}\bm{\partial}p + \nu\partial^2 \bm{u} \qquad \bm{\partial}\cdot \bm{u}=0
  \end{equation}
\noindent where $\bm{u}$ and $p$ are the velocity and pressure fields respectively, $\nu$ is the kinematic viscosity and $\rho_a$ the density of air (the list of all physical/chemical parameters is reported in the Tab.~\ref{tab:param}).
Instead of focusing on the evolution of the absolute humidity field, it is more convenient to model the supersaturation field $s=\mathrm{RH}-1$ (the exhaled air can be assumed to be saturated, or close to saturation \cite{morawska2009size}).
The supersaturation field is ruled by the  advection-diffusion equation \cite{celani2005droplet}
\begin{equation}
  \partial_t s + \bm{u}\cdot \bm{\partial} s= D_v \partial^2  s,
\label{eq-supersat}
\end{equation}
where $D_v$ is the water vapor diffusivity.
Eq.~(\ref{eq-supersat}) assumes that the saturated vapor pressure is constant, an assumption that holds true as long as the ambient is not much colder than the exhaled air, which is at about $30\,^oC$~\cite{morawska2009size}.

Eqs. (1) and (2) are solved within a domain box of length $L_x=4\,\mathrm{m}$, width $L_y=1.25\,\mathrm{m}$ and height $L_z=2.5\,\mathrm{m}$. The fluid is initially at rest, i.e. $\bm{u}(\bm{x},0)=\bm{0}$, and at the ambient supersaturation $s(\bm{x},0)=s_a=\mathrm{RH}_a-1$. 
Air is injected through a mouth opening located at $z_\mathrm{mouth}=1.6\,\mathrm{m}$ with area $A_\mathrm{mouth}=4.5\,\mathrm{cm^2}$ according to the time-varying profile representative of cough proposed by~\citet{gupta2009flow}.
The duration of the exhalation  is $0.4\,\mathrm{s}$ with a peak velocity of $13\,\mathrm{m/s}$. The Reynolds number (based on the peak velocity and on the mouth average radius) is about $9\times10^{3}$ and the resulting flow field is fully turbulent. The injected air is saturated, i.e.\ $s=0$. 
For the other domain boundaries, we prescribe the no-slip condition at the bottom ($z=0$) and left ($x=0$) boundaries and the free-slip condition at the top boundary ($z=L_z$), while applying the Dirichlet condition $s=s_a$.
For both the velocity and supersaturation field, we impose a convective outlet boundary condition at the right boundary ($x=L_x$). Finally, periodic boundary conditions apply at the side boundaries (i.e., $y=0$ and $y=L_y$).

\subsection{Lagrangian model for droplet transport and evaporation}
The exhaled droplets are modelled as an ensemble of $N$ particles dispersed within the airflow.
Droplets are initially at rest and their position is randomly distributed within a sphere of radius $1\,\mathrm{cm}$ inside the mouth opening.
Each droplet is ruled by the well-known set of equations \cite{maxey1983equation}
\begin{equation}
  \dot{\bm{X}_i}=\bm{U}_i(t)+\sqrt{2 D_v}\bm{\eta}_i (t)\qquad i=1, \ldots , N
  \label{eq:max1}
  \end{equation}
\begin{equation}
  \dot{\bm{U}_i}=\frac{\bm{u}(\bm{X}_i(t),t)-\bm{U}_i(t)}{\tau_i} +\bm{g}\qquad \tau_i =\frac{2 (\rho_{D\,i}/\rho_a) R_i^2(t)}{9\nu}
\label{eq:max2}
\end{equation}
where $N$ is the number of droplets,
$\bm{X}_i$ is the position of the $i$-th droplet and $\bm{U}_i$ its velocity, and $\bm{g}$ is the gravitational acceleration.
Each droplet is affected by a Brownian  contribution via the white-noise process $\bm{\eta}_i$.
Here, $\rho_{D\,i}$ is the density  of the i-th droplet.
Because the volume fraction of the liquid phase for cough is always smaller than $10^{-5}$ \cite{wang1993settling,bourouiba2014violent}, the back-reaction of the droplets to the flow can be safely neglected.
 Droplets are assumed to be made by salty water (water and NaCl) and a  solid insoluble part (mucus) \cite{vejerano2018physico}. Finally, $\tau_i $ is the Stokes relaxation time of the  droplet and $R_i$ is its radius.
Droplet radii evolve according to the dynamical equation~\cite{pruppacher1997microphysics} 
\begin{equation}
  \frac{d}{dt} R^2_i(t) = 2 C_R\left (1+s(\bm{X}_i(t),t)-e^{\frac{A}{R_i(t)}-B\frac{r_{N\,i}^3}{R_i^3(t)-r_{N\,i}^3 } }       \right )
\label{eq-radii}
\end{equation}
with the additional constraint for crystallization
\begin{equation}
  R_i(t)=r_{N\,i}  \qquad \mbox{for}\qquad s\le s_\mathrm{crh} \; \mbox{(crystallization).}
  \label{eq:R}
\end{equation}
No feedback effect to Eq.~(\ref{eq-supersat}) is considered here because of the very small values of the liquid volume fraction that we have specified above.
In Eq.~(\ref{eq-radii}), $C_R$ is the droplet condensational growth rate,
$s_\mathrm{crh}=\mathrm{CRH}-1$, where CRH is the  so-called crystallization RH (or efflorescence RH) for NaCl~\cite{lohmann2016introduction}. 
Refs.~\cite{biskos2006nanosize,zeng2014temperature} show
the weak dependence of CRH on temperature. $r_{N\,i} $ is the radius of the (dry) solid part of the i-th droplet when the salt is totally crystallized (i.e. below CRH). 
The dependence  of $r_{N\,i} $ on physical/chemical properties of the exhaled droplets is reported in the Supplementary Information together with the expressions of parameters $A$ and $B$. On the basis of the assumed parameters, the ratio $r_{N\,i}/R_i(0)$ is 0.16 which agrees well with the estimations discussed in  Ref.~\cite{nicas2005toward}.

\subsection{Numerical method and code implementation}

The flow solver is named \textit{Fujin} and is based on the finite-difference method for the spatial discretization and the (second-order) Adams-Bashfort scheme for the temporal discretization. The Poisson equation for the pressure is solved using the \texttt{2decomp} library coupled with a fast and efficient FFT-based approach.   The solver is parallelized using the MPI protocol and has been extensively validated in a variety of problems~\cite{rosti_brandt_2017a,rosti2019flowing,rosti_ge_jain_dodd_brandt_2019,rosti2020increase,olivieri2020dispersed}. See also: \texttt{https://groups.oist.jp/cffu/code}. The droplet dynamics is computed via Lagrangian particle tracking complemented by an established droplet condensation model that has been successfully employed in the past for the analysis of rain formation processes~\cite{celani2005droplet,celani2008equivalent,celani2009droplet}.
 Eqs.~(\ref{eq:max1}-\ref{eq:R}) for the droplet dynamics are here advanced in time using the explicit Euler scheme.

In the performed simulations, the domain is discretized with uniform spacing $\Delta x = 3.5 \, \mathrm{mm}$ in all directions, resulting in a total number of $N \approx 0.3$ billion grid points.
The results are first validated against the theoretical prediction for a turbulent puff~\cite{kovasznay1975unsteady} (see Fig.~\ref{fig:valPuff}). Moreover, we verified the convergence by comparing the results with those obtained by doubling the grid resolution. As shown in Fig.~\ref{fig:gridConv}, only minor differences are found both in terms of the probability density function of the particle evaporation time (Fig.~\ref{fig:gridConv}a) and of the cumulative number of droplets (Fig.~\ref{fig:gridConv}b).

\subsection{Scaling laws for a cloud of tracers in a puff}
By means of a simple phenomenological approach, we show how one can derive  the temporal scaling for the standard deviation of a cloud of tracers in a turbulent puff.  The starting point is the result obtained by~\citet{kovasznay1975unsteady} for the temporal scaling of the puff radius: $\sigma^v \sim t^{1/4}$ obtained by the author in terms of a simple eddy-viscosity approach. In order to determine the standard deviation, $\sigma$, for a cloud of tracers carried by the turbulent puff, one has to resort to the concept of relative dispersion. This latter can be described in terms of arguments {\it \`a la} Richardson~\cite{richardson1926atmospheric}. Accordingly, $\sigma (t) \sim \epsilon^{1/2} t^{3/2}$, where $\epsilon$ is the turbulence dissipation rate. By simple dimensional arguments,
\begin{equation}
  \epsilon (t) \sim \frac{\delta U^3}{\sigma^u}\qquad \delta U \sim \frac{\sigma^u}{t}
  \label{eq:epsilon}
  \end{equation}
from which one immediately gets: $\epsilon (t) \sim t^{-5/2}$.
The scaling law for $\epsilon$  immediately leads to the temporal scaling for the tracer cloud standard
deviation: $\sigma (t) \sim t^{1/4}$.  The reliability of this prediction is shown in Fig.~\ref{fig:valPuff}.

\subsection{Estimation of the viral load landing distance}
We estimate the distance from the mouth reached by the airborne droplets in the absence of external flows (depicted in Fig.~\ref{fig:air}f) as follows. First, we evaluate the settling velocity from Fig.~\ref{fig:air}e which clearly shows (when observed in linear scale) a linearly decreasing height of the viral load center of mass. From the same figure, we also obtain the time needed for the center of mass to reach the ground, $t_\mathrm{floor}$. We now split the airborne droplets in two groups, those that are inside the puff and those outside. For the former, we estimate the streamwise coordinate of their center of mass, $x_\mathrm{floor}$, as
\begin{equation}
  x_\mathrm{floor} - x_\mathrm{min}  = \int_{t_\mathrm{min}}^{t_\mathrm{floor}} v  \, \mathrm{d} t =\int_{t_\mathrm{min}}^{t_\mathrm{floor}} c_1 \, t^{-3/4}  \, \mathrm{d} t = \left[  4 c_1 \, t^{1/4}  \right]_{t_\mathrm{min}}^{t_\mathrm{floor}} = 4 c_1 \, \left( t_\mathrm{floor}^{1/4} - t_\mathrm{min}^{1/4} \right),
\end{equation}
where $t_\mathrm{min}$ is equal to $45\,\mathrm{s}$ and corresponds to the maximum simulated time unaffected by boundary condition effects, 
$x_\mathrm{min}$ is the streamwise coordinate of the viral load center of mass at time $t_\mathrm{min}$ for the considered droplets,
$v$ is the mean streamwise velocity (reported in Fig.~\ref{fig:valPuff}) and $c_1 = 1/2.2$ is a prefactor found by fitting the decay of $v$ with $t^{-3/4}$.
For the droplets outside the puff, we suppose they settle without changing their streamwise coordinate, such that $x_\mathrm{floor}=x_\mathrm{min}$. Finally, the center of mass of the viral load of the entire cloud of droplets has been obtained as the (initial volume) weighted average of the centers of mass of the two groups of droplets.

\clearpage
\subsection{Acknowledgments}
M.E.R., S.O. and A.M. acknowledge the computational time provided by HPCI on the Oakbridge-CX cluster in the Information Technology Center, The University of Tokyo, under the grant hp200157 of the ``HPCI Urgent Call for Fighting against COVID-19'' and the computer time provided by the Scientific Computing section of Research Support Division at OIST. A.M. thanks the financial support from the Compagnia di San Paolo, project MINIERA no. I34I20000380007. 

\subsection{Author contributions}
A.M. and M.E.R. conceived the original idea and planned the research.
M.E.R. developed the code and performed the numerical simulations.
A.M. and M.C. collected the experimental parameters for the droplet evaporation model from literature.
All authors analyzed data and outlined the manuscript content.
A.M., S.O., M.E.R. and A.S. wrote the manuscript with feedback from all authors.

\subsection{Competing interests}
The authors declare no competing interests.

\subsection{Data availability}
All data supporting the plots shown in the manuscript are available from the authors upon reasonable request.

\subsection{Code availability}
Details of the code used for the numerical simulations are provided in the Methods section and references therein. 

\clearpage

\setcounter{table}{0}
\makeatletter 
\renewcommand{\thetable}{Supplementary \@arabic\c@table}
\makeatother

\setcounter{figure}{0}
\makeatletter 
\renewcommand{\thefigure}{Supplementary \@arabic\c@figure}
\makeatother

\begin{table}[h]
\caption{Physical/chemical parameters representative of expiratory events and adopted in the present investigation.}
\label{tab:param}
\begin{ruledtabular}
\begin{tabular}{lcc}
  Mean ambient temperature & $T$ & $25\,^{\circ}\mathrm{C}$\\
  Crystallization (or efflorescence) RH & $\mathrm{CRH}$ & $45\%$\\
  Density of liquid water & $\rho_w$ & $9.97\times 10^2\, \mathrm{kg/m^3}$\\
  Density of soluble aerosol part (NaCl) & $\rho_s$ & $2.2\times 10^3\, \mathrm{kg/m^3}$\\
  Density of insoluble aerosol part (mucus) & $\rho_u$ & $1.5\times 10^3\, \mathrm{kg/m^3}$\\
  Density of dry nucleus & $\rho_N$ & $1.97\times 10^3\, \mathrm{kg/m^3}$\\
  Mass fraction of soluble material (NaCl) w.r.t. the total dry nucleus & $\epsilon_m$ & $0.75$\\
  Mass fraction of dry nucleus w.r.t. the total droplet & ${\cal C}$ & $1 \, \%$\\
  Specific gas constant of water vapor & $R_v$ & $4.6\times 10^2\, \mathrm{J/(kg\, K)}$\\
  Diffusivity of water vapor & $D_v$ & $2.5\times 10^{-5}\, \mathrm{m^2/s}$\\
  Density of air & $\rho_a$ & $1.18\, \mathrm{kg/m^3}$\\
  Kinematic viscosity of air & $\nu$ & $1.8\times 10^{-5}\, \mathrm{m^2/s}$\\
  Heat conductivity of dry air & $k_a$ & $2.6\times 10^{-2}\, \mathrm{W/K\, m}$\\
Latent heat for evaporation of liquid water & $L_w$ & $2.3\times 10^6\, \mathrm{J/kg}$ \\
Saturation vapor pressure & $e_\mathit{sat}$ & $0.616\, \mathrm{kPa}$\\
Droplet condensational growth rate & $C_R$ & $1.5  \times 10^{-10} \,\mathrm{m^2/s}$\\
Surface tension between moist air and salty water & $\sigma_w$ & $7.6\times 10^{-2}\, \mathrm{J/m^2}$\\
Molar mass of NaCl & $M_s$ & $5.9\times 10^{-2}\, \mathrm{kg/mol}$\\
Molar mass of water & $M_w$ & $1.8\times 10^{-2}\, \mathrm{kg/mol}$\\
Vertical distance of the mouth opening from the floor & $y_\mathrm{mouth}$ & $1.6\, \mathrm{m}$\\
Cross-sectional area of the mouth opening & $A_\mathrm{mouth}$ & $4.5\, \mathrm{cm^2}$\\
\end{tabular}
\end{ruledtabular}
\end{table}

\clearpage


\clearpage

\begin{figure}
    \centering
    \subfigure[a][]{
    \includegraphics[width=0.45\textwidth]{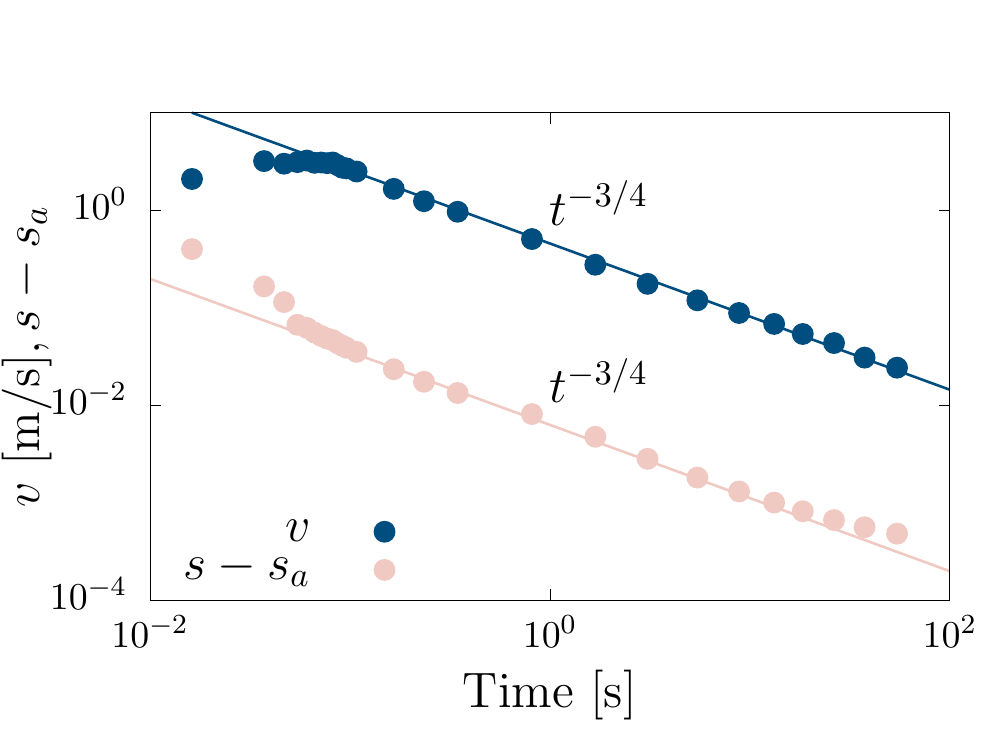}}
    \subfigure[b][]{
    \includegraphics[width=0.45\textwidth]{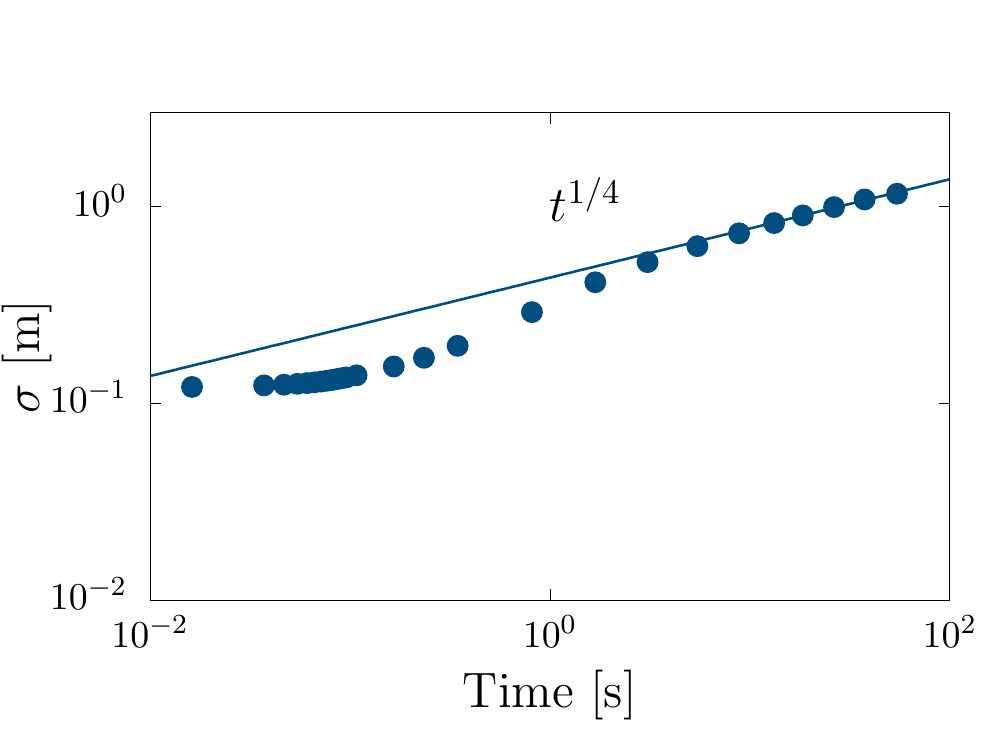}}
    \caption{\textbf{Validation of turbulent puff dynamics and tracers.} 
    \textbf{a} Time history of the mean streamwise velocity component $v$ (blue) and of the supersaturation field $s-s_a$ (pink).
 The symbols refer to the results from our direct numerical simulations.
The lines show the expected scaling for both the velocity and supersaturation field~\cite{kovasznay1975unsteady}.
\textbf{b} The standard deviation of a cloud of tracers as a function of time.  Symbols refer to the results from our direct numerical simulations; the continuous line is the expected scaling law obtained on the basis of the phenomenological arguments reported in Methods.}
    \label{fig:valPuff}
\end{figure}

\clearpage

\begin{figure}
    \centering
    \subfigure[a][]{
    \includegraphics[width=0.45\textwidth]{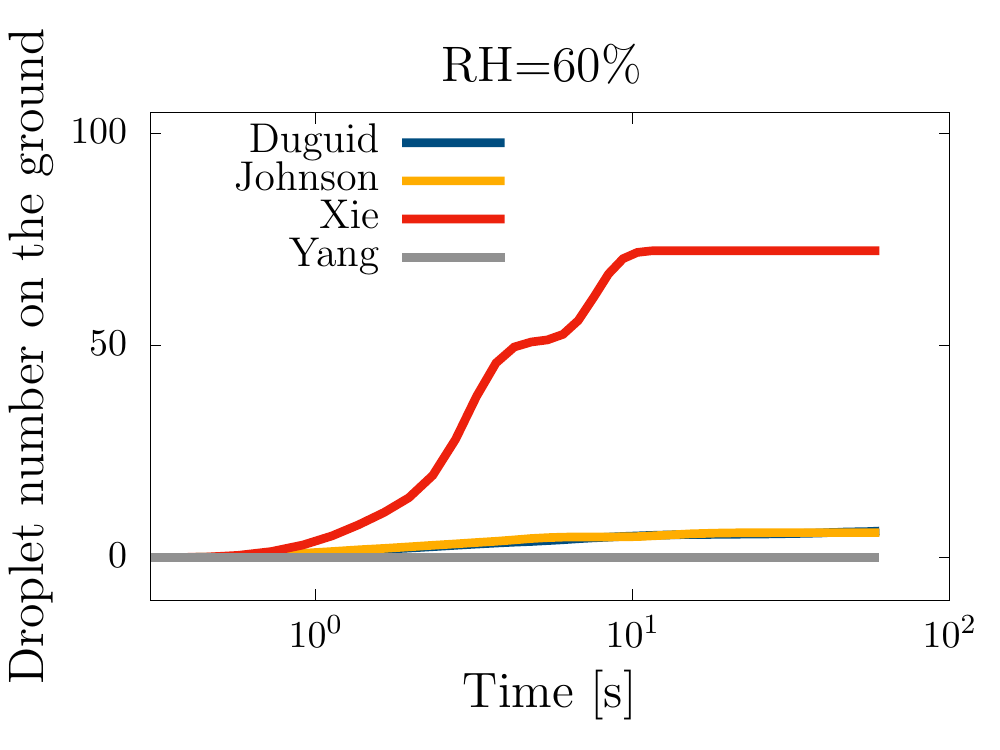}}
    \subfigure[b][]{
    \includegraphics[width=0.45\textwidth]{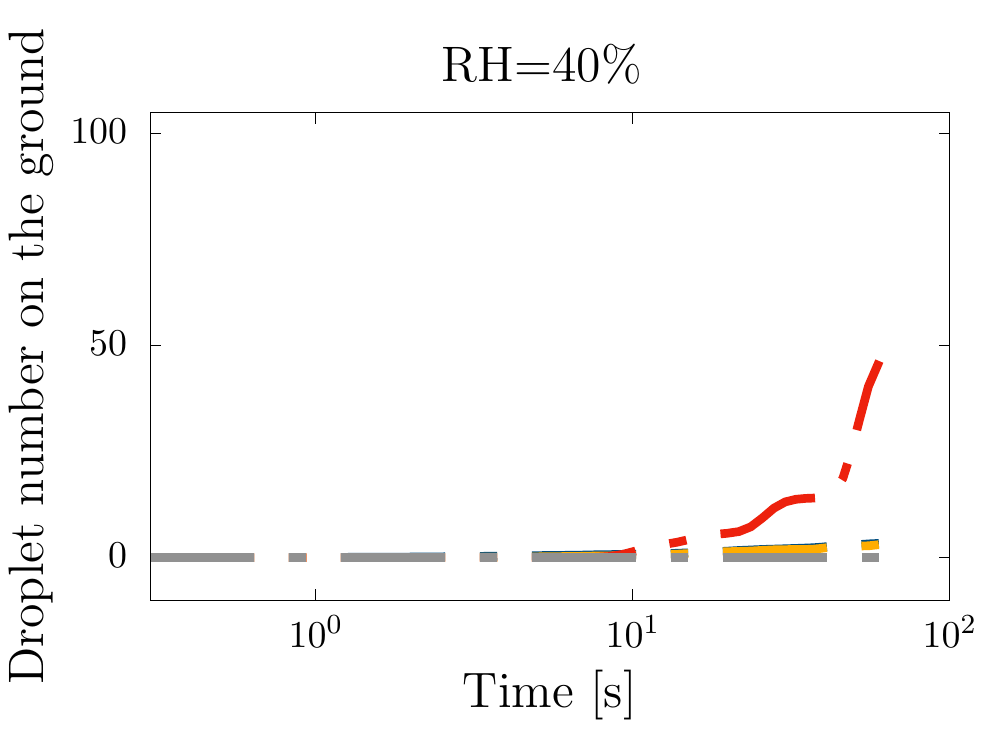}}
    \subfigure[c][]{
   \includegraphics[width=0.45\textwidth]{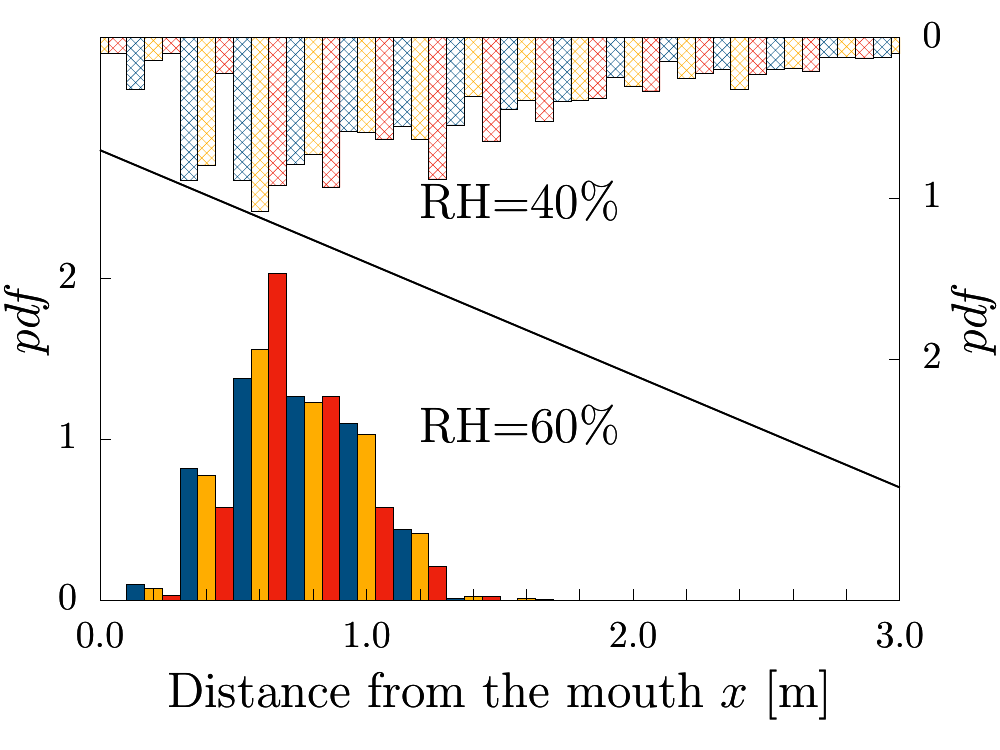}}
   \caption{\textbf{Sedimentation of large droplets.}
\textbf{a} Normalized number of droplets settling to the ground, obtained with the four different initial droplet size distributions proposed by~\citet{duguid1946size} (blue),~\citet{johnson2011modality} (yellow),~\citet{xie2009exhaled} (red) and~\citet{yang2007size} (gray). Here, the ambient relative humidity is RH=60$\%$. 
\textbf{b} Same as (a), for a dry environment, RH=40$\%$.
 \textbf{c} Probability density function of the distance from the mouth when droplets reach the ground obtained for all droplet size distributions; environmental relative humidity $\mathrm{RH}=60\%$ (solid) and $\mathrm{RH}=40\%$ (patterned).  Note that, no droplets sediment with the distribution by Yang.}
\label{fig:ext-sed}
\end{figure}

\clearpage


\clearpage

\begin{figure}
    \centering
    \includegraphics[width=0.65\textwidth]{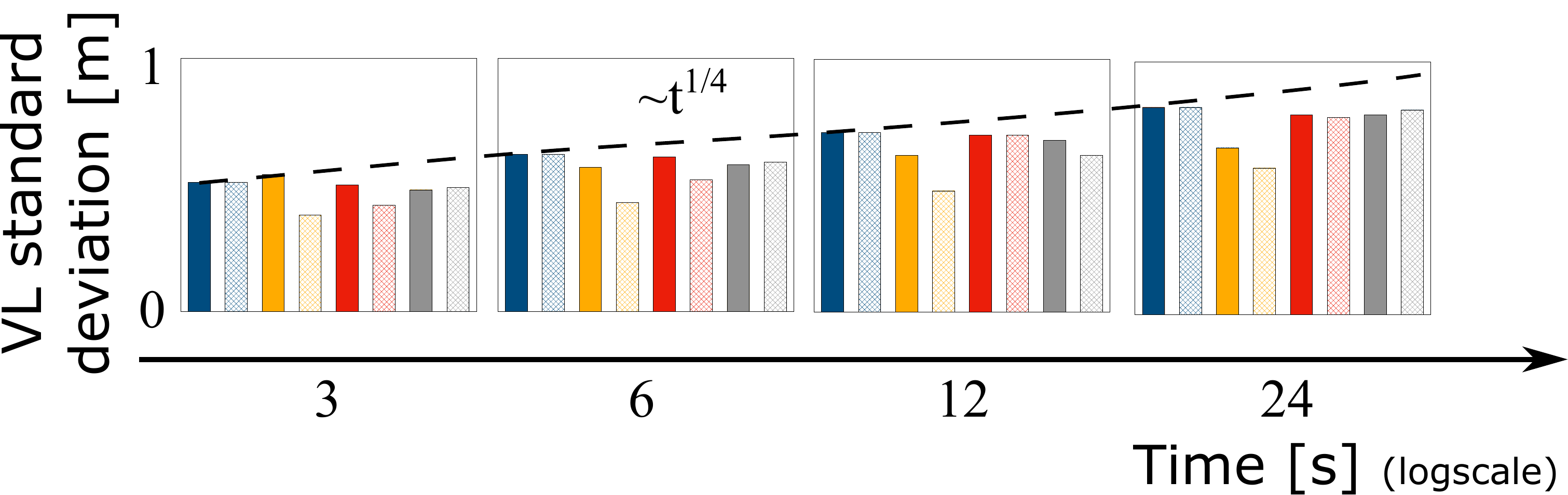} 
    \caption{
    \textbf{Airborne-transmitted droplets.}
    Time evolution of the viral load standard deviation for the eight numerical experiments performed. The dashed line represents the expected power-law growth predicted by means of phenomenological arguments (see Methods). The variability observed for the standard deviations associated to different initial droplet size distributions reaches values of about $30\%$.
     }
    \label{fig:sigma2}
\end{figure}

\clearpage

\begin{figure}
    \centering
    \subfigure[a][]{
    \includegraphics[width=0.45\textwidth]{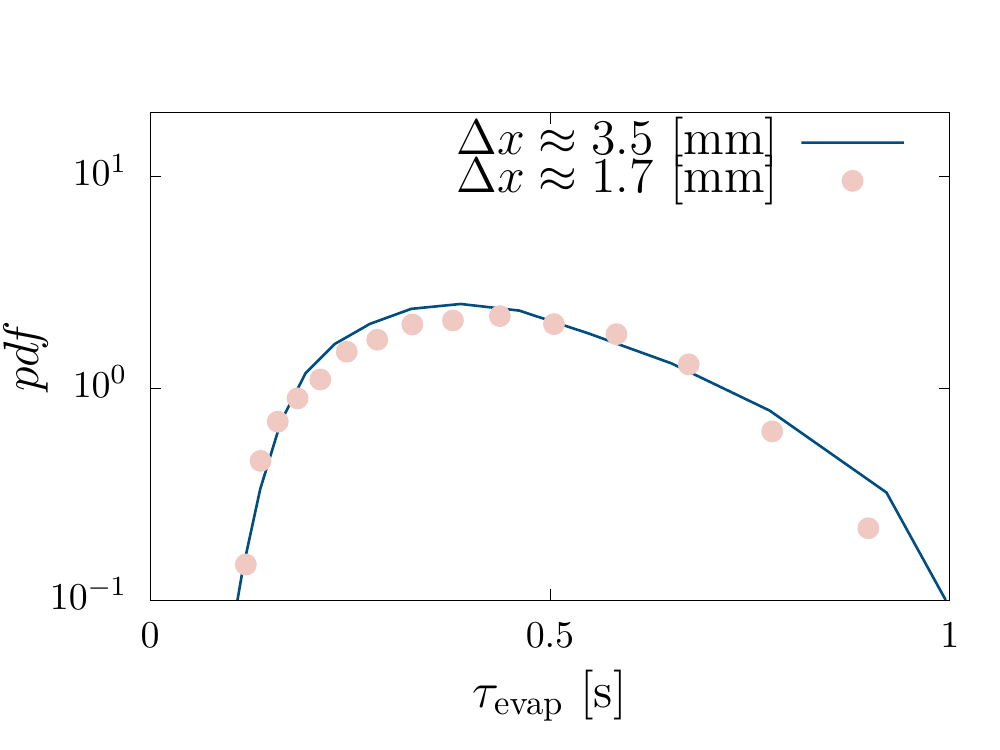}}
    \subfigure[b][]{
    \includegraphics[width=0.45\textwidth]{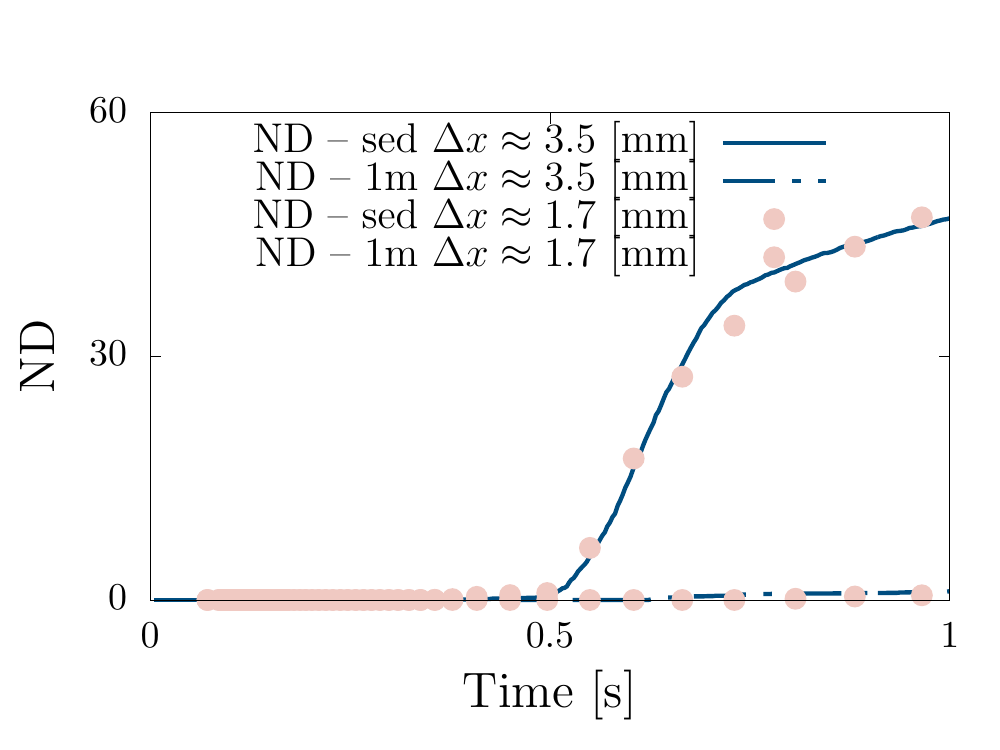}}
    \caption{\textbf{Grid convergence analysis.}
    Comparison between results computed with grid spacing $\Delta x=3.5\,\mathrm{mm}$ (blue lines) and $\Delta x=1.75\,\mathrm{mm}$ (pink symbols). 
     \textbf{a} Probability density function of the particle evaporation time.
     \textbf{b} Cumulative number of sedimenting and airborne droplets (in percentage with respect to the total number of droplets).}
    \label{fig:gridConv}
\end{figure}

%

\clearpage

\section{Supplementary information}


The complete list of physical and chemical parameters involved in our model is reported in Table~\ref{tab:param} along with their baseline values adopted in this investigation.
The dry nucleus of droplets is assumed to be composed by a soluble phase (NaCl) and a insoluble phase (mucus).
Given the typical value of the mass fraction for the former, the overall density of the dry nucleus can be expressed as 
\begin{equation}
\rho_N  = \frac{\rho_u}{1-\epsilon_m[1-(\rho_u/\rho_s)]}
\end{equation}
while the density of the entire $i$-th droplet is
\begin{equation}
\rho_{D\,i}  = \rho_w + (\rho_N-\rho_w)\left (\frac{r_{N\,i}}{R_i(t)}\right )^3,
\end{equation}
where the radius of the (dry) solid part of the droplet when NaCl is totally crystallized (i.e. below CRH) is given by
\begin{equation}
r_{N\,i}  = R_i(0)\left (\frac{{\cal C}\;\rho_w}{{\cal C}\;\rho_w +\rho_N(1-{\cal C})}\right )^{1/3}.
\end{equation}
Some additional expressions (which can be found in the literature~\cite{monteith2013principles,pruppacher1997microphysics} or derived by simple arguments) complete the description:
\begin{equation}
e_\mathit{sat}= 6.1078\times 10^2\,e^{(17.27\, T/(T+237.3))} \, \mathrm{Pa},
\label{eq:_esat}
\end{equation}
\begin{equation}
C_R=\left[\frac{\rho_w\, R_v\, (273.15+T)}{e_{sat}\,D_v}+\frac{\rho_w\,L_w^2}{k_a\,R_v\,(273.15+T)^2}-\frac{\rho_w\,L_w}{k_a(273.15+T)}\right ]^{-1},
\label{eq:CR}
\end{equation}
\begin{equation}
A=\frac{2 \sigma_w}{R_v (T+273.15) \rho_w}, 
\label{eq:A}
\end{equation}
\begin{equation}
B=\frac{n_s \Phi_s \epsilon_v M_w \rho_s}{M_s \rho_w}.
\end{equation}
Here, $n_s=2$ is the total number of ions into which a salt molecule dissociates, $\Phi_s=1.2$ is the practical osmotic coefficient of the salt in solution~\cite{liu2017short} and $\epsilon_v= \epsilon_m(\rho_N/\rho_s)$ is the volume fraction of dry nucleus with respect to the total droplet.
Finally, note that in Eqs.~\eqref{eq:_esat}, \eqref{eq:CR} and \eqref{eq:A} the temperature $T$ is expressed in degrees Celsius.

\end{document}